%% file: main.tex
\documentclass[aps,twocolumn,amsmath,amsfonts,nofootinbib,preprintnumbers,superscriptaddress,secnumarabic]{revtex4-1}
\usepackage[utf8]{inputenc}
\usepackage{bbold,graphicx,hyperref,amssymb,slashed,xspace}
\usepackage[dvipsnames]{xcolor}
\usepackage[section]{placeins}
\hypersetup{colorlinks=true,linkcolor=ForestGreen,citecolor=Maroon,filecolor=magenta,urlcolor=cyan}

\newcommand\sss{\scriptscriptstyle}
\newcommand{\mgamc}{{\sc MG5aMC}\xspace}
\newcommand{\mk}{{\sc MadFKS}\xspace}
\newcommand{\ml}{{\sc MadLoop}\xspace}
\newcommand{\ct}{{\sc CutTools}\xspace}
\newcommand{\nin}{{\sc Ninja}\xspace}
\newcommand{\coll}{{\sc Collier}\xspace}

\begin{document}
\title{Precision predictions for exotic lepton production at the Large Hadron Collider}

\author{A.~H.~Ajjath}
\email{aabdulhameed@lpthe.jussieu.fr}
\affiliation{Sorbonne Universit\'e, CNRS, Laboratoire de Physique Th\'eorique et Hautes \'Energies, LPTHE, F-75005 Paris, France}
\author{Benjamin~Fuks}
\email{fuks@lpthe.jussieu.fr}
\affiliation{Sorbonne Universit\'e, CNRS, Laboratoire de Physique Th\'eorique et Hautes \'Energies, LPTHE, F-75005 Paris, France}
\author{Hua-Sheng~Shao}
\email{huasheng.shao@lpthe.jussieu.fr}
\affiliation{Sorbonne Universit\'e, CNRS, Laboratoire de Physique Th\'eorique et Hautes \'Energies, LPTHE, F-75005 Paris, France}
\author{Yehudi~Simon}
\email{ysimon@lpthe.jussieu.fr}
\affiliation{Sorbonne Universit\'e, CNRS, Laboratoire de Physique Th\'eorique et Hautes \'Energies, LPTHE, F-75005 Paris, France}

\begin{abstract}
We calculate total and differential cross sections  for the pair production, at the Large Hadron Collider, of exotic leptons that could emerge from models with vector-like leptons and in Type-III seesaw scenarios. Our predictions include next-to-leading-order QCD corrections, and we subsequently match them with either parton showers, or threshold resummation at the next-to-next-to-leading logarithmic accuracy. Our results show an important increase of the cross sections relative to the leading-order predictions, exhibit a distortion of the shapes for various differential distributions, and feature a significant reduction of the scale uncertainties. Our predictions have been obtained from new {\sc FeynRules} model implementations and associated Universal {\sc FeynRules} Output (UFO) model libraries. This completes the set of next-to-leading-order implementations of new physics models featuring extra leptons that are publicly available on the {\sc FeynRules} model database.
\end{abstract}

\maketitle

\section{Introduction}

Several extensions of the Standard Model (SM) predict the existence of new exotic heavy leptons. These arise in particular in composite models~\cite{Panico:2015jxa, Cacciapaglia:2020kgq, Cacciapaglia:2022zwt}, grand unified theories~\cite{Morais:2020odg, Morais:2020ypd}, supersymmetric models~\cite{Martin:2009bg, Endo:2011mc, Araz:2018uyi}, left-right symmetric models~\cite{Pati:1974yy, Mohapatra:1974gc, Senjanovic:1975rk, Mohapatra:1977mj}, dark matter scenarios~\cite{Halverson:2014nwa}  or in Type~I~\cite{Minkowski:1977sc, Mohapatra:1979ia, Yanagida:1979as, Gell-Mann:1979vob, Glashow:1979nm, Shrock:1980ct, Schechter:1980gr, Cai:2017mow} and Type-III~\cite{Foot:1988aq, Cai:2017mow} seesaw models. In composite scenarios, the new physics particle spectrum features vector-like leptons (VLLs) transforming as electroweak $SU(2)_L$ singlets or doublets. In contrast, in seesaw and left-right models, neutrino masses are generated from Yukawa interactions of new electroweak singlets or triplets of fermions with the SM Higgs field and lepton weak doublet. Consequently, searches for new heavy leptons consist of an important component of the experimental beyond the SM (BSM) search programme at the Large Hadron Collider (LHC). 

The ATLAS and CMS collaborations have explored the associated parameter spaces, both for promptly-decaying~\cite{ATLAS:2018dcj, ATLAS:2019isd, ATLAS:2019kpx, ATLAS:2020wop, ATLAS:2021wob, ATLAS:2022yhd, ATLAS:2023, CMS:2017ybg, CMS:2018iaf, CMS:2018agk, CMS:2018jxx, CMS:2018iye, CMS:2019hsm, CMS:2019lwf, CMS:2021dzb, CMS:2022usq, CMS:2022cpe, CMS:2022nty} and long-lived~\cite{ATLAS:2019kpx, ATLAS:2022atq} extra leptons, and for a variety of mass ranges. Limits have been set on the mixing properties of long-lived heavy charge-neutral leptons for masses of 3--15~GeV, while short-lived neutral and charged leptons must have masses of at least about 950~GeV in Type-III seesaw models (for branching ratios of 1 in the final state considered). Moreover, in left-right models neutral lepton masses must be larger than 3~TeV for $W_R$ boson masses smaller than 4--5~TeV, and indirect probes for heavy neutrinos via vector-boson-fusion processes additionally constrain masses ranging up to 20~TeV for large mixings with the SM leptons and neutrinos. Finally, bounds on VLLs strongly depend on their representation under the electroweak symmetry group and on the VLL couplings to SM leptons. For instance, whereas weak doublets of VLL coupling to tau leptons are constrained to be heavier than about 1~TeV~\cite{ATLAS:2023}, the limit drops in the 100 -- 200 GeV range in the singlet case~\cite{CMS:2022nty}. On the other hand, LEP bounds on light VLLs are still relevant, and they impose a lower limit on the VLL mass. Such a limit lies in the 100~GeV range at best, depending on the details of the model~\cite{ALEPH:1996akm, DELPHI:2004ili, L3:2001xsz, OPAL:2002wkp}.

From the theoretical side, total and differential cross section calculations for collider processes involving extra neutrinos are known at next-to-leading order (NLO) in QCD in a generic simplified model describing the dynamics of the heavy neutrinos~\cite{Degrande:2016aje,Fuks:2020att}, for an effective left-right-symmetric scenario~\cite{Mattelaer:2016ynf} and for the production of type~III seesaw leptons~\cite{Ruiz:2015zca}. In addition, NLO-QCD predictions matched with threshold resummation at the next-to-leading-logarithmic (NLL) accuracy~\cite{Debove:2010kf,Fuks:2012qx,Fuks:2013vua} and approximate next-to-next-to-leading-order cross section matched with next-to-next-to-leading-logarithmic (NNLL) threshold resummation~\cite{Fiaschi:2020udf} can also be obtained from electroweakino pair-production processes in the Minimal Supersymmetric Standard Model, after decoupling all supersymmetric states but the produced electroweakinos. Beside differential and total rates, precision collider simulations in which NLO calculations are matched with parton showers (PS) are available for heavy neutrino simplified models~\cite{Degrande:2016aje, Mattelaer:2016ynf, Fuks:2020att}, as well as for VLL models for well-defined mixings between the composite and the SM sectors~\cite{Bhattiprolu:2019vdu}. However, simulations for Type-III seesaw scenarios and generic composite VLL setups are only available at leading order (LO) so far.

The goal of this paper is to fill this gap, and to report about the development of two new publicly available UFO~\cite{Degrande:2011ua} model libraries allowing for event generation at the NLO+PS accuracy. Our implementation are suitable in particular to describe VLL and Type-III seesaw lepton production processes from computations achieved by means of the precision Monte Carlo event generator {\sc MadGraph5\_aMC@NLO}~\cite{Alwall:2014hca, Frederix:2018nkq} (\mgamc). Moreover, we take the opportunity to update predictions for the total rates and the invariant mass distributions of the BSM Drell-Yan-like processes inherent to the models considered, and present results at NLO in QCD matched with threshold resummation at NNLL, following the formalism of~\cite{Sterman:1986aj, Catani:1989ne, Vogt:2000ci}.

The rest of this paper is organised as follows. In section~\ref{sec:theory}, we introduce two effective theoretical frameworks suitable for our calculations, a first one dedicated to a simplified VLL model and a second one to Type-III seesaw scenarios, and we report details about their implementation.In section~\ref{resum}, we briefly describe the formalism that we use to resum the large threshold logarithms. In section~\ref{sec:lhc}, we make use of the \mgamc platform to study the phenomenology of the two models at the NLO+PS accuracy, as well as of an in-house programme to handle predictions at NLO+NNLL in the strong coupling. We compute total rates for the production of extra leptons (section~\ref{sec:totalxsec}), invariant-mass distributions (section~\ref{sec:dsdm}), and we additionally present for the first time NLO+PS-accurate differential distributions relevant for experimental searches for VLLs and Type-III seesaw fermions (section~\ref{sec:dsdpt}). We summarise our work and conclude in section~\ref{sec:conclusion}.

\section{Theoretical models and the implementations}\label{sec:theory}

In order to study the phenomenology of the considered models, we construct two effective frameworks, one for each of the models. We minimally extend the SM in terms of fields and interactions, so that the resulting new physics parameter spaces are of small dimensionality. 

\subsection{A simplified model for VLL phenomenology}\label{sec:vll_theory}
We begin by considering an extension of the SM in which the theory field content includes a set of VLL fields. They are all colour singlet, but each of them lies in a different $SU(2)_L$ representation. They are correspondingly assigned different hypercharge $U(1)_Y$ quantum numbers. In order to be as model independent as possible, we adopt a simplified model approach and focus on extra leptons that are either electrically neutral or with an electric charge $Q=\pm 1$. These leptons are organised in (vector-like) $SU(2)_L$ doublets and singlets,
\begin{equation}\renewcommand{\arraystretch}{1.5}
  L^0 = \begin{pmatrix} N^0\\E^0 \end{pmatrix}\, , \qquad
  \tilde E^0\, , \qquad
  \tilde N^0\, ,
\label{eq:vllgauge}\end{equation}
where in this notation the superscript `0' indicates that the fields are gauge eigenstates, and the tilde above a field indicates that it is an $SU(2)_L$ singlet. We next implement the mixing between the SM and the new lepton fields. To this aim, we introduce an effective parametrisation apt to capture the main phenomenological features of the vector-like fields in a model-independent way, following guidelines introduced for vector-like quark setups~\cite{Buchkremer:2013bha, Fuks:2016ftf}. 

In practice, we assume that the mixing between the SM fields and the new leptons is small, so that the gauge interactions of the different fields are unaffected at the first order. Moreover, we implement the off-diagonal interactions of the exotic leptons with the SM ones through generic free parameters, which further open the VLL decay channels into a SM lepton and an electroweak boson. The corresponding Lagrangian, given in terms of mass eigenstates (the superscript `0' being therefore dropped), reads
\begin{equation}\label{eq:vlllag}\renewcommand{\arraystretch}{1.0}\begin{split}
    & {\cal L}_{\rm VLL} =  {\cal L}_{\rm SM} 
      + i \bar L\slashed{D} L - m_N \bar N N - m_E \bar E E\\ 
    & \ \
      + i \bar{\tilde{N}} \slashed{\partial} \tilde{N} - m_{\tilde{N}} \bar{\tilde{N}} \tilde{N}
      + i \bar{\tilde{E}}\slashed{D} \tilde{E} - m_{\tilde{E}} \bar{\tilde{E}} \tilde{E}\\
    & \ \
      + \sum_{\Psi=E,\tilde E}\! \Big[ h \bar \Psi \Big(\hat\kappa_{\sss L}^{\sss \Psi} P_L + \hat\kappa_{\sss R}^{\sss \Psi} P_R\Big) \ell 
      + \! \frac{g}{\sqrt{2}} \bar \Psi \slashed{W}^-\!\! \kappa_{\sss L}^{\sss \Psi} P_L \nu_\ell\\
    & \qquad\qquad 
      + \frac{g}{2 c_{\sss W}} \bar \Psi \slashed{Z} \Big(\tilde\kappa_{\sss L}^{\sss \Psi}P_L + \tilde\kappa_{\sss R}^{\sss \Psi} P_R\Big) \ell + {\rm H.c.} \Big] \\
    & \ \
      + \sum_{\Psi=N,\tilde N} \Big[ h \bar \Psi \hat\kappa_{\sss L}^{\sss \Psi} P_L \nu_\ell 
      + \frac{g}{2 c_{\sss W}} \bar \Psi \slashed{Z} \tilde\kappa_{\sss L}^{\sss \Psi}P_L \nu_\ell\\
    & \qquad\qquad 
      + \frac{g}{\sqrt{2}} \bar \Psi \slashed{W}^+ \Big(\kappa_{\sss L}^{\sss \Psi} P_L + \kappa_{\sss R}^{\sss \Psi} P_R\Big) \ell + {\rm H.c.} \Big]\,,
\end{split}\end{equation}
where ${\cal L}_{\rm SM}$ is the SM Lagrangian, and the parameters $m_N$, $m_E$, $m_{\tilde{N}}$ and $m_{\tilde{E}}$ stand for the masses of the four new fields in the physical basis, assuming that the masses of the doublet component fields can be different after electroweak symmetry breaking and particle mixing. The first two lines in this Lagrangian include, additionally to the SM Lagrangian, all gauge-invariant kinetic and mass terms for the new states. The gauge-covariant derivative operator $D_\mu$ is defined, for a generic field $\psi$, by
\begin{equation}
    D_\mu\psi = \partial_\mu\psi - i g' B_\mu Y \psi - i g W_\mu^k T^k \psi \,.
\end{equation}
Here, the coupling constants $g$ and $g'$ respectively stand for the weak $SU(2)_L$ and hypercharge $U(1)_Y$ coupling constants, and $B_\mu$ and $W_\mu^k$ are the associated gauge fields. The action of the hypercharge operator $Y$ on the field $\psi$ can be deduced from table~\ref{tab:vll_gaugefields}, as the representation to adopt for the $SU(2)$ generators $T^k$. In particular, $T^k=0$ for an $SU(2)_L$ singlet, and $T^k=\sigma^k/2$ for an $SU(2)_L$ doublet, with $\sigma^k$ being the Pauli matrices. In other words, we approximate mass eigenstates by gauge eigenstates in all kinetic terms, {\it i.e.}\ $L^0\approx L$, $\tilde E^0\approx \tilde E$ and $\tilde N^0\approx \tilde N$. 

\begin{table}[!t]
  \renewcommand{\arraystretch}{1.4}
  \setlength\tabcolsep{6pt}
  \begin{tabular}{c c c c}
    Field & Spin & Representation & Name\\ \hline
    $L^0$ & $(1/2,1/2)$ & $({\bf 1}, {\bf 2})_{-1/2}$ & {\tt VLL0}\\
    $\tilde N^0$ & $(1/2,1/2)$ & $({\bf 1}, {\bf 1})_{0\phantom{-/2}}$ & {\tt VLN0}\\
    $\tilde E^0$ & $(1/2,1/2)$ & $({\bf 1}, {\bf 1})_{-1\phantom{/2}}$ & {\tt VLE0}\\
  \end{tabular}
  \caption{Gauge eigenstates complementing the SM field content, their spin given as their representation under the $SO(1,3)$ group (second column), their $\text{SU}(3)_c\times \text{SU}(2)_L \times \text{U}(1)_Y$ (third column) representation and their name in the {\sc FeynRules} implementation (last column).} \label{tab:vll_gaugefields}
  \end{table}

The last four lines of the Lagrangian~\eqref{eq:vlllag} collect the effective interactions of each of the four VLLs considered with a SM lepton ($\ell$ standing for the charged lepton field and $\nu_\ell$ for the neutrino one), and either the Higgs boson $h$, the $W$ boson or the $Z$ boson. In \eqref{eq:vlllag}, all flavour indices are understood so that each of the $\kappa$, $\hat\kappa$ and $\tilde \kappa$ couplings has to be seen as a vector in the flavour space. Moreover, $c_{\sss W}$ refers to the cosine of the electroweak mixing angle.

In the following, the exact values of the $\kappa$, $\hat\kappa$ and $\tilde \kappa$ coupling vectors are irrelevant, provided that they are not too large to guarantee that the VLL states have a narrow width, and not too small so that they can promptly decay into a lepton+electroweak boson system within LHC detector scales. In a hadron collision process in which the new leptons are pair produced, such couplings indeed only appear in the heavy particle decays.
  
\begin{table}[!t]
  \renewcommand{\arraystretch}{1.4}
  \setlength\tabcolsep{6pt}
  \begin{tabular}{c c c c c c}
  Field & Spin & Name & PDG & Mass & Width \\ \hline
  $N$ & $(1/2,1/2)$& {\tt VLLN} & 9000001 & {\tt MVLLN} & {\tt WVLLN}\\
  $E$ & $(1/2,1/2)$ & {\tt VLLE} & 9000002 & {\tt MVLLE} & {\tt WVLLE}\\
  $\tilde N$ & $(1/2,1/2)$ & {\tt VLN} & 9000003 & {\tt MVLN} & {\tt WVLN}\\
  $\tilde E$ & $(1/2,1/2)$ & {\tt VLE} & 9000004 & {\tt MVLE} & {\tt WVLE}\\
  \end{tabular}
  \caption{Mass eigenstates supplementing the SM, with their spin quantum number (second column), name used in the {\sc FeynRules} convention (third column) and adopted PDG identifier (fourth column). In the last two columns, we provide the {\sc FeynRules} symbols associated with the particle masses and widths.} \label{tab:vll_massfields}
\end{table}

In order to allow for phenomenological studies of the model, we implement it in the {\sc FeynRules} package~\cite{Christensen:2009jx,Alloul:2013bka}, starting from the SM implementation that is shipped with the programme. We include the definitions of the gauge eigenstates of~\eqref{eq:vllgauge}, together with the corresponding mass eigenstates appearing in the Lagrangian~\eqref{eq:vlllag}. Information on these fields, their names in the {\sc FeynRules} conventions, and the Particle Data Group (PDG) identifiers that we have adopted for the physical fields, are provided in tables~\ref{tab:vll_gaugefields} and \ref{tab:vll_massfields}. These tables also include the symbols associated with the mass and width of the physical fields. All BSM couplings appearing in~\eqref{eq:vlllag} have been implemented as three-vectors in the flavour space, following the convention of table~\ref{tab:vll_params}. This table also includes information on the Les Houches block structure used to organise all model external parameters~\cite{Skands:2003cj}, as required by all high-energy physics programmes relying on {\sc FeynRules} for model implementation. Moreover, their specific contributions to any process can be turned off through a dedicated interaction order named {\tt VLL} (see the {\sc FeynRules} manual~\cite{Alloul:2013bka}).

\begin{table}
  \renewcommand{\arraystretch}{1.4}
  \setlength\tabcolsep{7pt}
  \begin{tabular}{c c c c}
    Couplings & Names & Les Houches blocks\\\hline
    $(\hat\kappa_{\sss L}^{\sss E})_i$, $(\hat\kappa_{\sss R}^{\sss E})_i$  
       & {\tt KLLEH[i]}, {\tt KRLEH[i]} & {\tt KLLEH}, {\tt KRLEH}\\
    $(\hat\kappa_{\sss L}^{\sss {\tilde E}})_i$, $(\hat\kappa_{\sss R}^{\sss {\tilde E}})_i$
       & {\tt KLEH[i]}, {\tt KREH[i]} & {\tt KLEH}, {\tt KREH}\\
    $(\hat\kappa_{\sss L}^{\sss N})_i$ & {\tt KLLNH[i]} & {\tt KLLNH}\\
    $(\hat\kappa_{\sss L}^{\sss {\tilde N}})_i$ & {\tt KLNH[i]} & {\tt KLNH}\\\hline
    $(\tilde\kappa_{\sss L}^{\sss E})_i$, $(\tilde\kappa_{\sss R}^{\sss E})_i$  
       & {\tt KLLEZ[i]}, {\tt KRLEZ[i]} & {\tt KLLEZ}, {\tt KRLEZ}\\
    $(\tilde\kappa_{\sss L}^{\sss {\tilde E}})_i$, $(\tilde\kappa_{\sss R}^{\sss {\tilde E}})_i$
       & {\tt KLEZ[i]}, {\tt KREZ[i]} & {\tt KLEZ}, {\tt KREZ}\\
    $(\tilde\kappa_{\sss L}^{\sss N})_i$ & {\tt KLLNZ[i]} & {\tt KLLNZ}\\
    $(\tilde\kappa_{\sss L}^{\sss {\tilde N}})_i$ & {\tt KLNZ[i]} & {\tt KLNZ}\\\hline
    $(\kappa_{\sss L}^{\sss E})_i$  & {\tt KLLEW[i]} & {\tt KLLEW}\\
    $(\kappa_{\sss L}^{\sss {\tilde E}})_i$ & {\tt KLEW[i]} & {\tt KLEW}\\
    $(\kappa_{\sss L}^{\sss N})_i$, $(\kappa_{\sss R}^{\sss N})_i$ 
       & {\tt KLLNW[i]}, {\tt KRLNW[i]} & {\tt KLLNW}, {\tt KRLNW}\\
    $(\kappa_{\sss L}^{\sss {\tilde N}})_i$, $(\kappa_{\sss R}^{\sss {\tilde N}})_i$ 
       & {\tt KLNW[i]}, {\tt KRNW[i]} & {\tt KLNW}, {\tt KRNW}\\
  \end{tabular}
  \caption{Three-point VLL coupling strengths to a SM lepton and an electroweak boson, given together with the associated {\sc FeynRules} symbol and the corresponding Les Houches block. The indice $i$ denotes a generation index ranging from $1$ to $3$.} \label{tab:vll_params}
\end{table}

\subsection{An effective Type-III seesaw Lagrangian}\label{sec:typeiii_theory}
In Type-III seesaw models, neutrino masses are generated through the interactions of the SM Higgs field $\Phi$ with the SM leptons and at least two generations of extra fermions lying in the adjoint representation of $SU(2)_L$ and with zero hypercharge. In the following, we make use of two-component Weyl fermion notation for all fields, and omit all SM and BSM generation indices for clarity. In such a formalism, the Lagrangian of the model is expressed in terms of the SM weak doublet of left-handed leptons $L_L$, the SM weak singlet of right-handed charged leptons $E_R^c$ ($E_R$ being thus the corresponding left-handed Weyl spinor), and the weak triplet of extra lepton $\Sigma^k$ (with $k=1, 2, 3$ being an $SU(2)_L$ adjoint index). The three gauge eigenstates $\Sigma^k$ can be conveniently related to states of definite electric charge $E^\pm \equiv \Sigma^\pm$ (of charge $Q=\pm 1$) and $N \equiv \Sigma^0$ (of charge $Q=0$) by introducing the matrix representation for the $SU(2)_L$ triplets $\Sigma^i{}_j$, with $i$ and $j$ referring to fundamental indices of $SU(2)_L$. We obtain,
\begin{equation}
  \Sigma^i{}_j = \frac{1}{\sqrt{2}} (\sigma^k)^i{}_j \Sigma^k = 
    \begin{pmatrix} \frac{1}{\sqrt{2}} N & E^+\\  E^- & -\frac{1}{\sqrt{2}} N\\ \end{pmatrix}\,.
\end{equation}

The Type-III Lagrangian is given by
\begin{equation}\label{eq:lag_type3}\begin{split}
 &  {\cal L}_{\rm TypeIII} = 
        \overline{{\cal L}}_{\rm SM} + {\cal L}_{\rm kin} 
    \\ & \hspace*{0.5cm} + \Big(
        y_\ell\ \Phi^\dag L_L.E_R   + 
        2 y_\Sigma\ \Phi\!\cdot\! \Big[\Sigma^k  T^k  L_L\Big] +
     {\rm H.c.} \Big)\,.
\end{split}\end{equation}
In our notation, $\overline{{\cal L}}_{\rm SM}$ stands for the reduced SM Lagrangian in which all terms involving a leptonic field have been removed, and ${\cal L}_{\rm kin}$ collects all gauge-invariant kinetic terms for the (two-component) leptonic fields $L_L$, $E_R$ and $\Sigma$, and a mass term for the $\Sigma$ field (of mass $m_\Sigma$). Moreover, the matrices $T^k =\sigma^k/2$ stand for the generators of $SU(2)_L$ in the fundamental representation, and the explicit scalar product appearing on the second line refers to the $SU(2)$-invariant product of two fields lying in its fundamental representation. In Type-III models, charged lepton and neutrino masses are driven by the SM leptonic $3\times 3$ Yukawa matrix $y_\ell$, and the heavy neutrino Yukawa matrix $y_\Sigma$ whose size depends on the number of generations of new fermions. 

From the LHC physics point of view, we can simplify the model presented above by emphasising the focus on the lightest of all $\Sigma$ states, that are assumed to be the only ones to which the LHC would be sensitive. This strategy follows that introduced in ref.~\cite{Biggio:2011ja}. Under such an assumption, the relevant part of the Yukawa matrix $y_\Sigma$ (in the $\Sigma$-flavour $\times$ SM-flavour space) becomes a vector in the SM-flavour space, that we take real for simplicity. 

The mass eigenstates of the model hence include three generations of physical up-type and down-type quarks, and four generations of physical charged leptons ($\ell'$) and Majorana neutrinos ($\nu'$). After electroweak symmetry breaking, the Lagrangian~\eqref{eq:lag_type3} induces a mixing between the three SM leptons and the new states $\Sigma$, rendering at least two neutrinos massive. Introducing the three $4\times 4$ mixing matrices in the lepton flavour space $U_L^\ell$, $U_R^\ell$ and $U^{\nu}$, lepton gauge and mass eigenstates are related by~\cite{Li:2009mw},
\begin{equation}\label{eq:typeiii_mixing}\begin{split}
  \begin{pmatrix}\ E_L\\ E^- \end{pmatrix} = U^\ell_L \ \ell'_L \,, \ 
  \begin{pmatrix}\ E_R\\ E^{+c} \end{pmatrix} = U^\ell_R\ \ell'_R  \,, \ 
  \begin{pmatrix}\ \nu_L\\ N \end{pmatrix} = U^\nu\ \nu' \,,
\end{split}\end{equation}
where the three-component vector in the flavour space $E_L$ ($\nu_L$) stands for the down-type (up-type) component of the weak doublet of left-handed SM leptons. As the new fermion masses of ${\cal O}(m_\Sigma)$ are expected to be heavy compared with the neutrino masses of ${\cal O}(y_\Sigma v)$, with $v$ being the SM Higgs vacuum expectation value, the three mixing matrices can be expanded at the first order in $yv/m_\Sigma$~\cite{Abada:2007ux,Abada:2008ea},
\begingroup\renewcommand*{\arraystretch}{1.5}\setlength\arraycolsep{3pt}
\begin{eqnarray}
  &&U_L^\ell = \begin{pmatrix}
    \mathbb{1}-\varepsilon & \frac{v}{m_\Sigma} y_\Sigma\\
    - \frac{v}{m_\Sigma} y_\Sigma^\dag & 1-\varepsilon'
  \end{pmatrix}\,,\ \
  U_R^\ell = \begin{pmatrix}
    \mathbb{1} & \frac{M_\ell v}{m^2_\Sigma} y_\Sigma\\
    - \frac{M_\ell v}{m^2_\Sigma} y_\Sigma^\dag & 1
  \end{pmatrix}\,,\nonumber \\
  && U^\nu = \begin{pmatrix}
    \big[\mathbb{1}-\frac12 \varepsilon\big] U_{\rm PMNS} & \frac{v}{\sqrt{2} m_\Sigma} y_\Sigma\\
    - \frac{v}{\sqrt{2} m_\Sigma} y_\Sigma^\dag & 1- \frac12 \varepsilon'
  \end{pmatrix}\,.
\end{eqnarray}
\endgroup
These expressions depend on the quantities $\varepsilon$ and $\varepsilon'$, that are $3\times 3$ and scalar objects in the SM flavour space respectively,
\begin{equation}
  \varepsilon = \frac{v^2}{2m_\Sigma^2}\ y_\Sigma y_\Sigma^\dag
  \quad\text{and}\quad
  \varepsilon' = \frac{v^2}{2m_\Sigma^2} y_\Sigma^\dag  y_\Sigma\,,
\end{equation}
as well as on the SM lepton mass matrix $M_\ell$ (that is diagonal in the flavour space) and on the unitary Pontecorvo-Maki-Nakagawa-Sakata (PMNS) matrix $U_{\rm PMNS}$. The latter can be defined from the neutrino oscillation parameters, namely the neutrino mixing angles $\theta_{ij}$ (with $i, j = 1, 2, 3$), the Dirac $CP$-violating phase $\varphi_{CP}$ and the two Majorana $CP$-violating phases $\varphi_1$ and $\varphi_2$.

\begin{table}
  \renewcommand{\arraystretch}{1.4}
  \setlength\tabcolsep{5pt}
  \begin{tabular}{c c c c c}
    Field      & Spin      & Representation                        & \# generations & Name\\ \hline
    $L_L$      & $(1/2,0)$ & $({\bf 1}, {\bf 2})_{-1/2}$           & $3$            & {\tt LLw}\\
    $E_R$      & $(1/2,0)$ & $({\bf 1}, {\bf 1})_{1\phantom{-/2}}$ & $3$            & {\tt ERw}\\
    $\Sigma^k$ & $(1/2,0)$ & $({\bf 1}, {\bf 3})_{0\phantom{-/2}}$ & $1$            & {\tt Sigw}\\
  \end{tabular}
  \caption{Gauge eigenstates associated with the leptonic sector of the Type-III seesaw model, their spin given as their representation under the $SO(1,3)$ group (second column), their $\text{SU}(3)_c\times \text{SU}(2)_L \times \text{U}(1)_Y$ representation (third column), and their name in the {\sc FeynRules} implementation (last column).}\label{tab:typeiii_gaugefields}
  \begin{tabular}{c c c c c c}
  Field    & Spin        & Name       & PDG     & Mass        & Width \\ \hline
  $e^-$    & $(1/2,1/2)$ & {\tt e}    & 11      & {\tt Me}     & --\\
  $\mu^-$  & $(1/2,1/2)$ & {\tt mu}   & 13      & {\tt MMu}    & --\\
  $\tau^-$ & $(1/2,1/2)$ & {\tt ta}   & 15      & {\tt MTA}    & --\\
  $E^-$    & $(1/2,1/2)$ & {\tt SigM} & 9000017 & {\tt MSigma} & {\tt WSigM}\\
  $\nu_1$  & $(1/2,1/2)$ & {\tt v1}   & 12      & {\tt Mv1}    & --\\
  $\nu_2$  & $(1/2,1/2)$ & {\tt v2}   & 14      & {\tt Mv2}    & --\\
  $\nu_3$  & $(1/2,1/2)$ & {\tt v3}   & 16      & {\tt Mv3}    & --\\
  $N$      & $(1/2,1/2)$ & {\tt Sig0} & 9000018 & {\tt MSigma} &  {\tt WSig0}\\
  \end{tabular}
  \caption{Mass eigenstates that either supplement the SM or whose definition is altered relatively to the SM, with their spin representation (second column), name used in the {\sc FeynRules} convention (third column) and adopted PDG identifier (fourth column). In the last two columns, we provide the {\sc FeynRules} symbols associated with the particle masses and widths.}\label{tab:typeiii_massfields}
\end{table}

We implement the Type-III model described above in the {\sc FeynRules} package~\cite{Christensen:2009jx,Alloul:2013bka} following the same method that has been used for the Type~II seesaw implementation~\cite{Fuks:2019clu}. We begin with the implementation of the SM shipped with {\sc FeynRules}, from which all lepton definitions and related Lagrangian terms have been modified.
In practice, we have modified all lepton and neutrino definitions so that two-component left-handed Weyl fermions~\cite{Duhr:2011se} are used instead for the gauge eigenstates of the model ($E_R^c$, $L_L$).
Next, we add definitions for the fermionic triplets $\Sigma^k$, still using left-handed Weyl fermions, and incorporate the mixing relations~\eqref{eq:typeiii_mixing} for the definition of the four physical charged lepton states and the four physical neutrino states in terms of all gauge eigenstates. Finally, we map the physical two-component fermions of the model into the corresponding Dirac fields (charged leptons) and Majorana fields (neutrinos). More information on all fields included in the model implementation is given in tables~\ref{tab:typeiii_gaugefields} and \ref{tab:typeiii_massfields} (representation, names in the {\sc FeynRules} conventions, PDG identifiers, symbols for masses and widths).

\begin{table}
  \renewcommand{\arraystretch}{1.4}
  \setlength\tabcolsep{7pt}
  \begin{tabular}{c c c c c}
    Parameter & Name & LH block & LH counter\\\hline
    \hline
    $(y_\Sigma)_e$    & {\tt ySigma[1]} & {\tt YSIGMA} & 1\\
    $(y_\Sigma)_\mu$  & {\tt ySigma[2]} & {\tt YSIGMA} & 2\\
    $(y_\Sigma)_\tau$ & {\tt ySigma[3]} & {\tt YSIGMA} & 3\\
    $m_\Sigma$        & {\tt MSigma}    & {\tt MASS} & 9000017\\
    \hline
    $m_{\nu_1}$       & {\tt Mv1}    & {\tt MASS} & 12\\
    $\Delta m^2_{21}$ & {\tt dmsq21} & {\tt MNU}  & 2\\
    $\Delta m^2_{31}$ & {\tt dmsq31} & {\tt MNU}  & 3\\
    \hline
    $\theta_{12}$ & {\tt th12} & {\tt PMNS}& 1\\
    $\theta_{23}$ & {\tt th23} & {\tt PMNS}& 2\\
    $\theta_{13}$ & {\tt th13} & {\tt PMNS}& 3\\
    $\varphi_{\rm CP}$ & {\tt delCP} & {\tt PMNS}& 4\\
    $\varphi_1$ & {\tt PhiM1} & {\tt PMNS}& 5\\
    $\varphi_2$ & {\tt PhiM2} & {\tt PMNS}& 6\\
  \end{tabular}
  \caption{External parameters defining the leptonic sector of the Type-III seesaw model, including the neutrino parameters in the context of a normal mass hierarchy (so that $m_{\nu_1} < m_{\nu_2} < m_{\nu_3}$). Each parameter is given together with the symbol used in the {\sc FeynRules} implementation, and the corresponding Les Houches (LH) block and counter information.} \label{tab:typeiii_params}
\end{table}

The new physics parameters $y_\Sigma$ and $m_\Sigma$ appearing in the Lagrangian~\eqref{eq:lag_type3} are implemented in a standard way, together with the neutrino oscillation parameters dictating the values of the PMNS matrix. Moreover, we assume a normal neutrino mass hierarchy and set the masses of the three lightest neutrinos $m_{\nu_1}$, $m_{\nu_2}$ and $m_{\nu_3}$ from the value of the smallest neutrino mass ($m_{\nu_1}$ in our case), and the neutrino squared mass differences $\Delta m_{21}^2$ and $\Delta m_{31}^2$,
\begin{equation}
    m_{\nu_2} = \sqrt{m_{\nu_1}^2 + \Delta m_{21}^2}\ \  \text{and}\ \ 
    m_{\nu_3} = \sqrt{m_{\nu_1}^2 + \Delta m_{31}^2}\,.
\end{equation}
More information on the free parameters of the leptonic/neutrino sector of the model is provided in table~\ref{tab:typeiii_params} ({\sc FeynRules} names and Les Houches block structure).

\subsection{From Lagrangian to events at the LHC}

In order to handle LHC simulations at NLO-QCD matched with PS, we make use of the two {\sc FeynRules} model implementations detailed in sections~\ref{sec:vll_theory} and \ref{sec:typeiii_theory}, and jointly use them with the {\sc MoGRe} package (version 1.1)~\cite{Frixione:2019fxg}, {\sc NloCt}~(version 1.0.1)~\cite{Degrande:2014vpa} and {\sc FeynArts}~(version 3.9)~\cite{Hahn:2000kx}. This allows us to renormalise the bare Lagrangians~\eqref{eq:vlllag} and \eqref{eq:lag_type3} relatively to $\mathcal{O}(\alpha_s)$ QCD interactions, and generate UFO model files~\cite{Degrande:2011ua} including both tree-level interactions, UV counterterms, and the so-called $R_2$ Feynman rules required for the numerical evaluation of the numerators of one-loop integrands in a four-dimensional spacetime. Such UFO models can subsequently be used with \mgamc~\cite{Alwall:2014hca,Frederix:2018nkq} for LO and NLO calculations in QCD, as well as by {\sc Herwig++}~\cite{Bellm:2015jjp} and {\sc Sherpa}~\cite{Gleisberg:2008ta} at LO.

Before closing this subsection, we provide information on the \mgamc\ framework~\cite{Alwall:2014hca, Frederix:2018nkq} which employ to carry out fixed-order (N)LO and (N)LO+PS calculations. \mgamc handles infrared singularities inherent to NLO calculations via the FKS method~\cite{Frixione:1995ms,Frixione:1997np}, in an automated way through the \mk module~\cite{Frederix:2009yq,Frederix:2016rdc}. The evaluation of UV-renormalised one-loop amplitudes is achieved by switching dynamically between several integral-reduction techniques that work either at the integrand level (like the OPP method~\cite{Ossola:2006us} or Laurent-series expansion~\cite{Mastrolia:2012bu}) or through tensor-integral reduction~\cite{Passarino:1978jh,Davydychev:1991va,Denner:2005nn}. This has been automated in the \ml\ module~\cite{Hirschi:2011pa,Alwall:2014hca}, that exploits the public codes \ct~\cite{Ossola:2007ax}, \nin~\cite{Peraro:2014cba,Hirschi:2016mdz} and \coll~\cite{Denner:2016kdg}. Moreover, one-loop computations have been optimised at the integrand level through an in-house procedure inspired by the idea of {\sc\small OpenLoops}~\cite{Cascioli:2011va}. Finally, NLO+PS predictions are obtained by matching fixed-order calculations with PS according to the {\sc MC@NLO} method~\cite{Frixione:2002ik}.

\section{Improving theory accuracy beyond NLO: threshold resummation}\label{resum}

For the Drell-Yan-like processes considered, it is well-known that large logarithms spoil the convergence of the perturbative series when the invariant mass $M$ of the final-state system approaches the hadronic centre-of-mass energy $\sqrt{s}$. This calls for a proper resummation of soft-gluon radiation, or at least a matching with PS as achieved in the \mgamc\ framework. In this section, we briefly describe, for the convenience of readers, the theoretical formalism that we adopt for fixed-order calculations matched with threshold resummation. Additional details and an extensive description can be found in section~2 of~\cite{Ajjath:2022kpv}.

At the partonic level, the corresponding kinematic region is defined in terms of the partonic scaling variable $z=M^2/\hat s$ when $z\to 1$, with $\sqrt{\hat s}$ being the partonic centre-of-mass energy. In this limit, the perturbative coefficients in the cross sections get contributions in the form of $\alpha_s \ln (1-z)$ originating from soft gluon emission, which could be of ${\cal O}(1)$ and thus potentially spoil the perturbative convergence of the usual series in $\alpha_s$. This issue can be resolved by reorganising the perturbative expansion in an alternative manner, and resumming the large logarithms in $\alpha_s \ln (1-z)$ to all orders in $\alpha_s$.

Resummation calculations are conveniently carried out in the Mellin $N$-space conjugate to $z$, where the $z\to 1$ limit thus corresponds to the large $N$ region. The resummed partonic cross section in the Mellin space, denoted by $\Delta_{q\bar{q}}^{\rm res}(N,M^2,\mu_F^2)$ with $\mu_F$ being the factorisation scale, is defined in eq.~(2.47) from~\cite{Ajjath:2022kpv} for a generic process with a colourless final state. For a Drell-Yan-like process and at the $k^{th}$ logarithmic accuracy (N$^k$LL), it reads
\begin{align}\label{DeltaN}
    & \Delta_{q\bar{q}}^{\rm res}(N,M^2,\mu_F^2) \Big|_{\rm {N^kLL}}  = \tilde{g}_{0,q\bar q}(M^2,\mu_F^2,\mu_R^2) \Big|_{\rm {N^kLO}}
    \nonumber \\
    &\quad \times 
    \exp \Big( g_{1,q\bar q} (\omega)\ln N + \sum_{j=2}^{k+1}{a_s^{j-2}(\mu_R^2)g_{j,q\bar q}(\omega)}\Big)\,.
\end{align}
In this expression, the $\tilde{g}_{0,q\bar q}|_{\rm {N^kLO}}$ factor collects the $N$-independent terms of the first $k+1$ coefficients of the usual  $a_s$ perturbative expansion (in Mellin space), where we have introduced the short-hand notation $a_s(\mu_R^2)=\alpha_s(\mu_R^2)/4\pi$ with $\mu_R$ denoting the renormalisation scale. This factor is process dependent, and it gets contributions from virtual corrections and soft real emission. In the exponent, the process-independent ({\it i.e.}\ universal) coefficients $g_{j,q\bar q} (\omega)$ with  $j>0$ receive contributions from the threshold logarithmic terms originating from real emission. Given $\omega = 2 \beta_0 a_s (\mu_R^2)\ln N\sim \mathcal{O}(1)$ with $\beta_0$ being the first coefficient of the QCD beta function, they effectively resum these logarithmic contributions to all orders in $\alpha_s$. We refer to appendix~\ref{app:resummationcoefficiebts} for the analytical form of the various coefficients appearing in~\eqref{DeltaN} and that are relevant for NLO+NNLL calculations for the Drell-Yan-like processes considered.

As the separation of the $N$-independent and $N$-dependent pieces in~\eqref{DeltaN} is not unambiguous, different resummation schemes have been proposed and described in section 2.4 of~\cite{Ajjath:2022kpv}. In the present paper, we consider the so-called ${\overline {\rm N}}_1$ resummation scheme for simplicity. 

Resummed calculations must then have to be matched with fixed-order predictions at (N)LO. This is achieved by adding the resummed and (N)LO results and subtracting of all double-counted contributions. The latter correspond to the (N)LO soft-virtual terms of the partonic cross section, which can be obtained by expanding $\Delta_{q\bar{q}}^{\rm res}(N,M^2,\mu_F^2)$ at $\mathcal{O}(\alpha_s^{b(+1)})$, where $b$ stands for the power in $\alpha_s$ of the LO contributions (that is 0 here).

\section{Cross sections for extra lepton production at the LHC}\label{sec:lhc}

In this section, we compute total and differential cross sections relevant for the production of additional leptons such as those appearing in the models introduced in section~\ref{sec:theory}. Predictions at (N)LO and (N)LO+PS are obtained within the \mgamc framework (version 3.3.0), using the UFO models developed in this work, and we employ {\sc Pythia}~8.2~\cite{Sjostrand:2014zea} to deal with the simulation of the QCD environment (parton showering and hadronisation). On the other hand, total rate calculations matching fixed-order predictions with soft-gluon resummation are derived with an in-house code.

We define the electroweak sector through three independent input parameters that we choose to be the $Z$-boson mass $m_Z$, the electromagnetic coupling constant evaluated at the $Z$-pole $\alpha(m_Z)$, and the Fermi constant $G_F$,
\begin{equation}\begin{split}
    & m_Z = 91.1876~{\rm GeV}\,, \quad
    \alpha^{-1}(m_Z) = 127.9\,,\\
    & G_F = 1.1663787 \cdot 10^{-5}~{\rm GeV}^{-2}\,.
\end{split}\end{equation}
In addition, the CKM matrix is taken diagonal, the pole mass of the top quark $m_t = 172.7$~GeV, and we consider $n_q=5$ active quark flavours. Moreover, the widths of all particles appearing in the relevant diagrams have been set to zero. Our predictions make use of the {\tt CT18NNLO}~\cite{Hou:2019efy} set of parton distribution functions (PDFs), which are provided by {\sc LHAPDF}~\cite{Buckley:2014ana} that we also use to control the renormalisation group running of the strong coupling $\alpha_s$. For predictions in the Type-III seesaw model, we safely set the elements of the $\varepsilon$ matrix to zero, as they turn to be negligible once bounds from flavour and electroweak precision data are accounted for~\cite{Biggio:2019eeo}.

The central value of the renormalisation and factorisation scales is set to the invariant mass $M$ of the produced di-lepton system, $\mu_R = \mu_F = M$. Scale uncertainties are then evaluated through the usual seven-point variation method, in which the renormalisation and factorisation scales are varied independently by a factor of two up and down relative to their central value with the two extreme cases $\mu_R/\mu_F=4$ or $1/4$ being excluded.

\subsection{Total cross sections at the LHC}\label{sec:totalxsec}

We dedicate this section to an overview of the behaviour of the total cross sections for exotic lepton production at the LHC with $\sqrt{s} = 14$~TeV, as a function of the lepton mass. We study the production of a pair of electrically-charged VLLs, and we consider both the cases of an $SU(2)_L$ singlet and doublet of VLLs,
\begin{equation}
    p p \to \tilde E^+ \tilde E^-\,, \qquad
    p p \to  E^+ E^-\,. 
\label{eq:process_vll}\end{equation}
Moreover, we also explore the production of a pair of singly-charged Type-III leptons that are mostly weak triplets,
\begin{equation}
    p p \to E^+ E^- \equiv \Sigma^+ \Sigma^-\,.
\label{eq:process_typeiii}\end{equation}
For the last process, we introduced the abusive notation $\Sigma^\pm \equiv E^\pm$ to make an explicit distinction between the VLLs appearing in the model of section~\ref{sec:vll_theory} (process~\eqref{eq:process_vll} and notation of table~\ref{tab:vll_massfields}), and those inherent to the Type-III seesaw model of section~\ref{sec:typeiii_theory} (process~\eqref{eq:process_typeiii} and notation of table~\ref{tab:typeiii_massfields}). We do not consider any other pair-production mechanism ({\it i.e.}\ the production of a pair of neutral or doubly-charged leptons, or of an associated pair of leptons of different charges), as cross section predictions are not expected to exhibit a fundamentally different behaviour due to the purely electroweak nature of the processes involved. We indeed focus, in the following, on the impact of higher-order corrections that only depends on the quark/gluon nature of the initial state. Our analysis therefore equally applies to charged-current and neutral-current production processes, the only difference between the various channels being the normalisation of the (differential) rates. However, as our UFO model files are public and \mgamc\ is a general-purpose event generator, interested readers can study by themselves any other process, both at (N)LO and (N)LO+PS.

\begin{figure}
    \centering\includegraphics[width=0.48\textwidth]{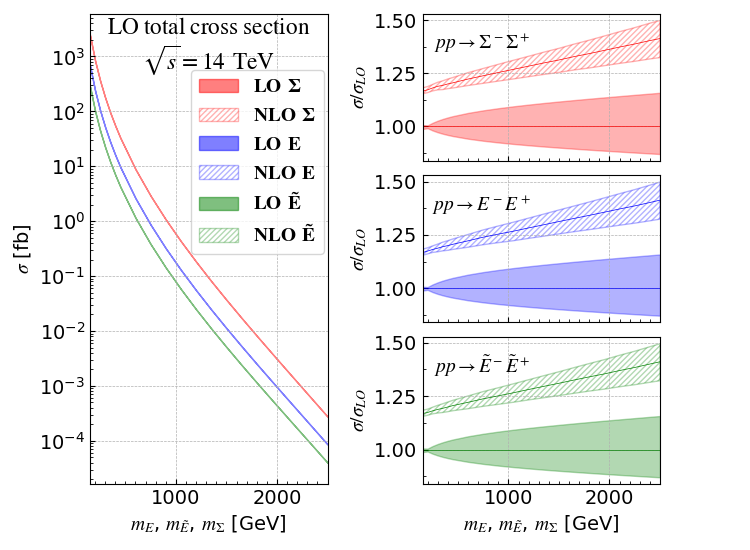}
    \caption{Total cross sections for the production of charged leptons typical from VLL and Type-III seesaw models, presented as a function of the lepton mass. We consider $SU(2)_L$ singlet (green), doublet (blue) and triplet (red) leptons, and the LHC at 14~TeV. Predictions are shown at LO (left panel), as well as in the form of LO and NLO $K$-factors together with the associated scale uncertainties (right panel).}\label{fig:NLO_LO_Xsec}
\end{figure}

In the left panel of figure~\ref{fig:NLO_LO_Xsec}, we report LO production cross sections for the three processes of eqs.~\eqref{eq:process_vll} and \eqref{eq:process_typeiii}. The cross sections are found to span about 7 orders of magnitude for exotic lepton masses varying from 200~GeV to 2.5~TeV. Cross sections around 100--1000~fb are found for small lepton masses of a few hundreds of GeV, whereas the production rates drop to the 0.001--1~fb regime for leptons of 1--2~TeV, making the potential observation of such BSM particles at the LHC more challenging. Moreover, for a given lepton mass, the production of a pair of weak-triplet states ($pp\to \Sigma^+ \Sigma^-$; red) is favoured over that of weak-doublet states ($p p \to  E^+ E^-$; blue), while the latter is favoured over the production of weak-singlet states ($p p \to  \tilde E^+ \tilde E^-$; green). Such a hierarchy, as well as the relative differences observed between the rates that are factors of a few, can be understood from the different $SU(2)_L$ representations of the fields considered, together with the Drell-Yan-like nature of the lepton pair-production mechanism.

In the right panel of figure~\ref{fig:NLO_LO_Xsec}, we present the corresponding $K$-factors, that we define, for a given lepton mass, as the ratio of a cross section to the associated LO one at central scale. $K$-factors are shown both at LO (shaded area) and NLO (hatched area), together with the associated scale uncertainties. We observe mild $K$-factor values at NLO, which vary in the 1.15--1.40 range as a function of the exotic lepton mass. Moreover, the $K$-factors are found to be (almost) independent of the process. Such a result is not surprising as the underlying Born contribution factorises in the case of a Drell-Yan-like process. Uncertainties are significantly larger at LO than at NLO, and vary in the last case from a few percents at small lepton masses to about 10\% for larger masses. This behaviour stems from the typically larger invariant masses associated with heavier di-lepton systems, that naturally enhance the importance of the threshold logarithms that ought to be resummed.

\begin{figure}
    \centering\includegraphics[width=0.48\textwidth]{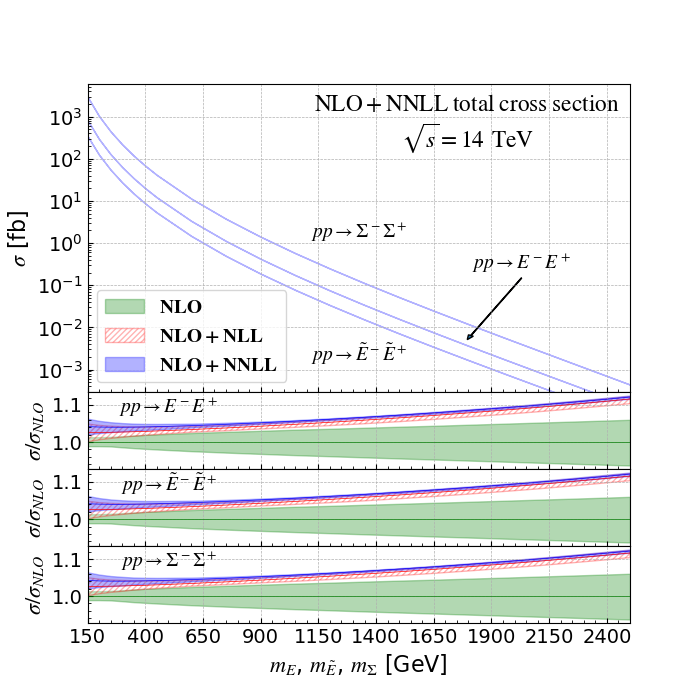}
    \caption{Total NLO+NNLL cross sections (upper panel) for the production of vector-like and Type-III seesaw leptons, presented as a function of the mass of the exotic lepton and for the LHC at a centre-of-mass energy of 14~TeV. We also display the ratios of the NLO (green), NLO+NLL (red) and NLO+NNLL (blue) rates to the NLO ones (with a central scale choice), together with the associated scale uncertainties.}\label{fig:NNLL_Xsec}
\end{figure}

In the upper panel of figure~\ref{fig:NNLL_Xsec}, we show the production rates obtained after matching NLO predictions with soft gluon resummation at the NNLL accuracy, these results being currently the best theoretical predictions of the total cross sections for the processes considered. In order to estimate the associated impact, we present, in the three lower panels of figure~\ref{fig:NNLL_Xsec}, the ratio of the NLO, NLO+NLL and NLO+NNLL rates to the NLO one for the three processes. By virtue of the factorisation properties of the Born contributions, the predictions for these ratios are mostly independent of the process. We observe a mild increase of the total rate once threshold resummation is included, although this increase is mostly driven by the leading and next-to-leading logarithmic contributions. NNLL contributions indeed barely modify the total rates. Numerically, these enhancements with respect to the NLO predictions are found to be about 5\% for scenarios with light leptons, and range to about 10\% for scenarios featuring heavier leptons. In addition, it is observed that the resummed results (at NLO+NLL and NLO+NNLL) lie outside the error bands of the NLO ones when a lepton mass of a few hundreds GeV is considered. This situation is similar to the case of the SM Drell-Yan process. 

As typical from resummation calculations, the most notable feature of the NLO+NNLL rate concerns the drastic reduction of the scale uncertainties inherent to the predictions, that are reduced below the percent level in average, regardless of the lepton mass. The necessity of incorporating soft gluon resummation in the predictions is made even more evident for scenarios featuring heavy exotic leptons. Here, the scale uncertainty bands at NLO+(N)NLL drastically decrease, which could be anticipated as we deal with a perturbative calculations in which $\alpha_s$ is even smaller (the scale at which it is evaluated being trivially larger). This essentially cures the counter-intuitive large scale uncertainties observed at NLO.

\begin{figure*}
    \centering
    \includegraphics[width=0.48\textwidth]{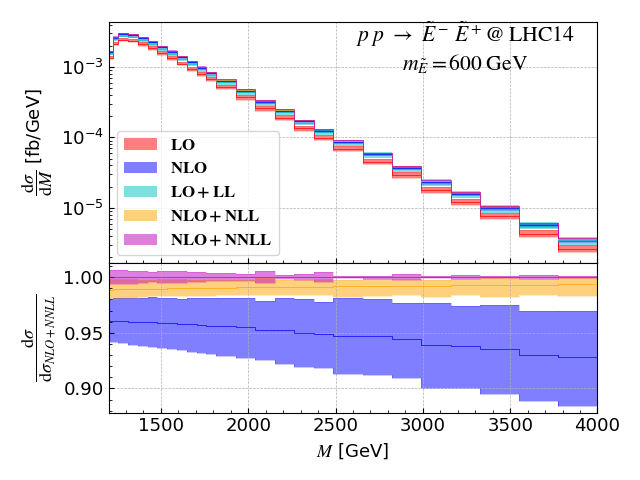}
    \includegraphics[width=0.48\textwidth]{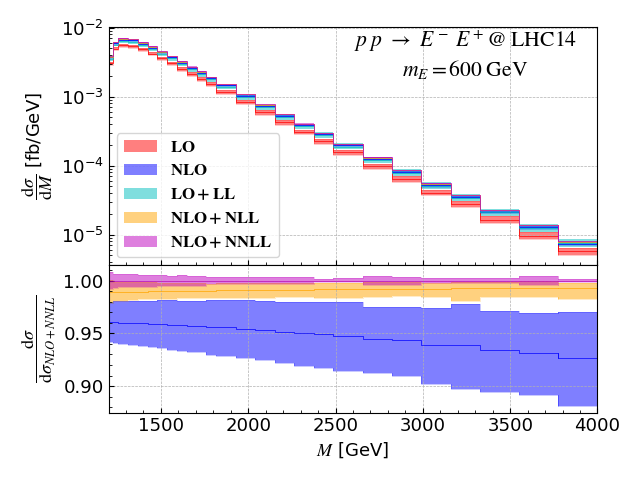}
    \includegraphics[width=0.48\textwidth]{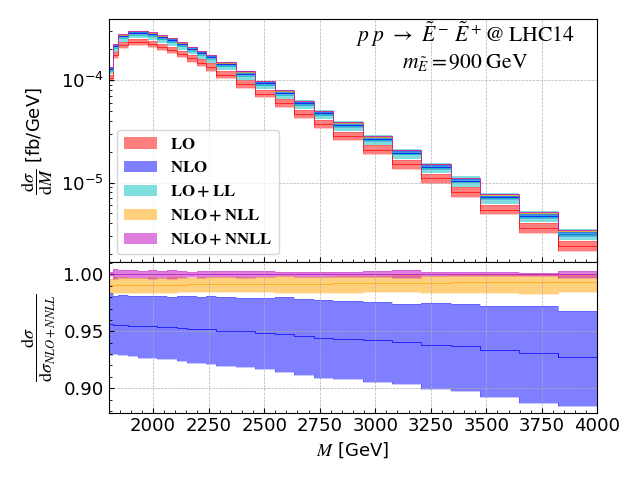}
    \includegraphics[width=0.48\textwidth]{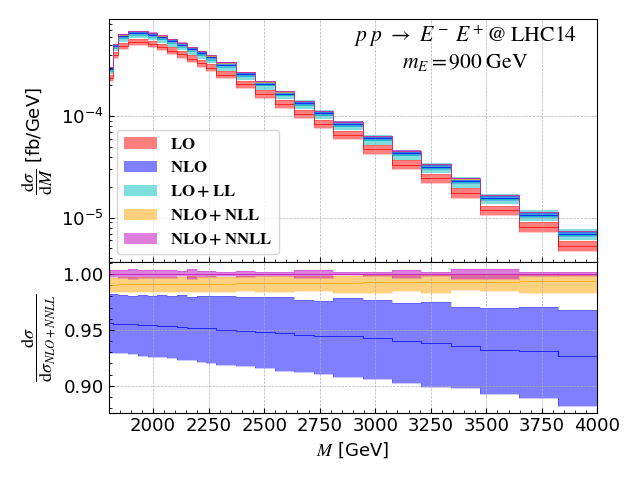}
    \includegraphics[width=0.48\textwidth]{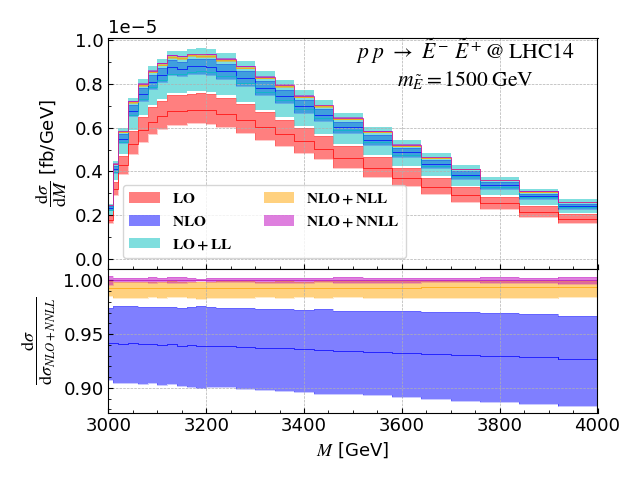}
    \includegraphics[width=0.48\textwidth]{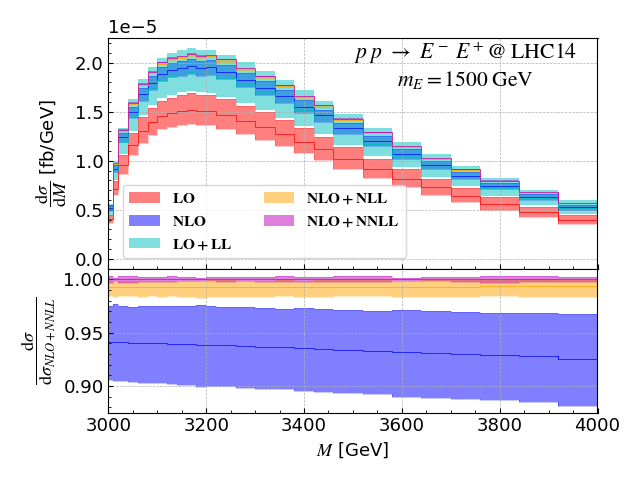}
    \caption{Invariant mass spectrum for the processes~\eqref{eq:process_vll}, together with the associated scale uncertainties (upper inset of each panel). We show fixed-order results at LO (red) and NLO (blue), as well as after matching them with threshold resummation at  LO+LL (cyan), NLO+NLL (yellow) and NLO+NNLL (purple), and we consider new lepton masses of 600 GeV (top row), 900~GeV (middle row) and 1.5~TeV (bottom row). We additionally present the bin-by-bin ratios of the NLO, NLO+NLL and NLO+NNLL spectra to the NLO+NNLL ones, with the associated uncertainties (lower inset in each panel).} \label{fig:dsdM_vll}
\end{figure*}

\subsection{Invariant mass distributions}\label{sec:dsdm}

\begin{figure}[!t]
    \centering
    \includegraphics[width=0.48\textwidth]{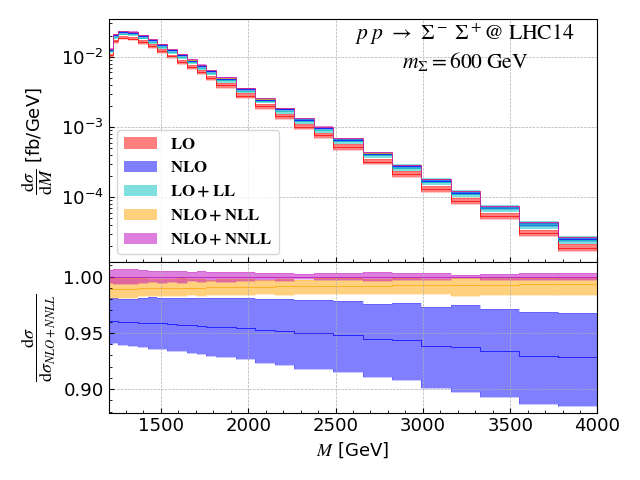}
    \includegraphics[width=0.48\textwidth]{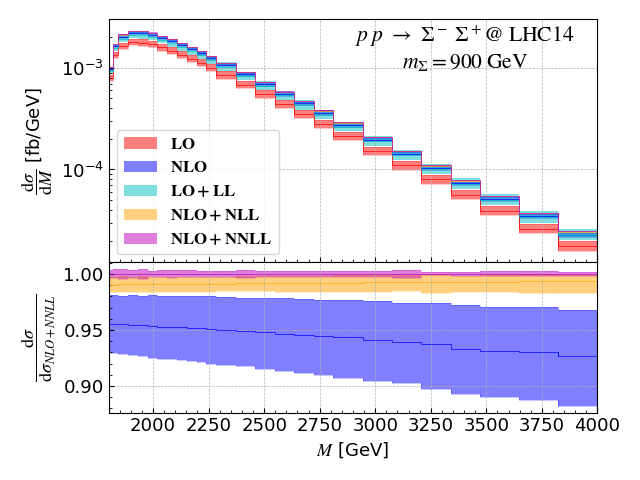}
    \includegraphics[width=0.48\textwidth]{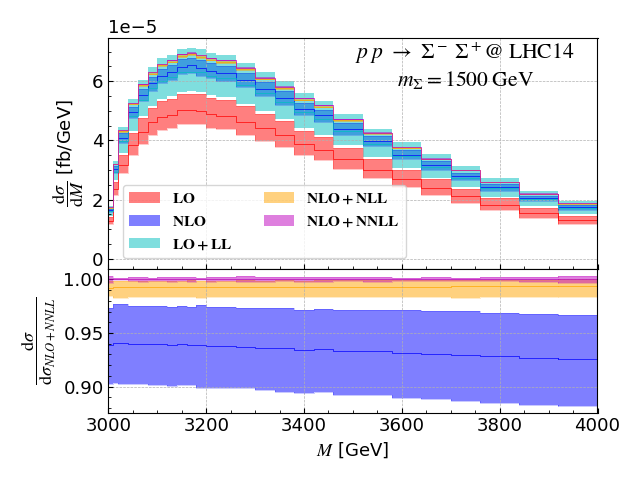}
    \caption{Same as figure~\ref{fig:dsdM_vll} but for Type-III seesaw leptons and the process~\eqref{eq:process_typeiii}.}\label{fig:dsdM_typeiii}
\end{figure}

In this section, we consider again the processes~\eqref{eq:process_vll} and \eqref{eq:process_typeiii}, and we calculate the associated invariant-mass distributions ${\rm d}\sigma/{\rm d}M$, with $M$ standing for the invariant mass of the di-lepton system. As an illustration, we choose three typical exotic lepton masses of 600~GeV, 900~GeV and 1.5~TeV, which correspond to three scenarios that have not been fully excluded yet by the LHC experiments. The definition of these scenarios is driven by current experimental search results. The CMS collaboration has indeed observed an excess of $2.8\sigma$ when searching for VLLs with a mass of 600~GeV~\cite{CMS:2022cpe}, whereas VLLs of 900 GeV can already be probed by existing run~2 CMS and ATLAS searches. In contrast, extra leptons with a mass of 1500 GeV lie beyond the current reach, but they could potentially be probed at future LHC runs. In addition, all chosen extra lepton masses lead to PDF uncertainties under fair control~\cite{Frixione:2019fxg}. The results are shown in figures~\ref{fig:dsdM_vll} and \ref{fig:dsdM_typeiii}.

In the upper inset in each plot, we display fixed-order predictions and the associated scale uncertainties at LO (red) and NLO (blue), as well as after matching them with threshold resummation at LO+LL (cyan), NLO+NLL (yellow) and NLO+NNLL (purple). We have checked that for invariant mass distributions, (N)LO+PS results coincide with (N)LO calculations since this observable is insensitive to PS effects. (N)LO+PS predictions are therefore not included in the figures. In general, we observe a rapid increase of the differential cross section close to the kinematic threshold $M_0$ equals to twice the heavy lepton mass, until it peaks at $M\approx 1.06 M_0$ before falling off toward higher invariant-mass values.

For any given $M$ value, fixed-order LO predictions (red) are found to be notably smaller. Their matching with predictions including the resummation of the LL contributions (cyan) yields a significant enhancement of the central value. It reaches 15\%--25\% in the peak region, and 30\% at larger invariant masses where the relevant phase space region is closer to the partonic threshold ($z\to 1$). Resummation effects are therefore more prominent in the latter case. Nevertheless, scale uncertainties in both cases remain large, and the two sets of predictions do not overlap within their error bands. This effect can be tamed down after including NLO corrections. In this case, both NLO+NLL and NLO+NNLL predictions are found to agree with each other once uncertainties are accounted for, whereas NLO spectra are slightly smaller for all considered $M$ values.

\begin{figure*}
    \centering
    \includegraphics[width=0.48\textwidth]{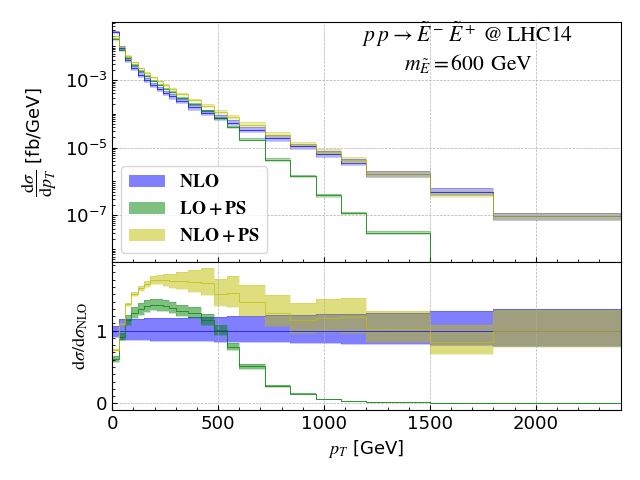}
    \includegraphics[width=0.48\textwidth]{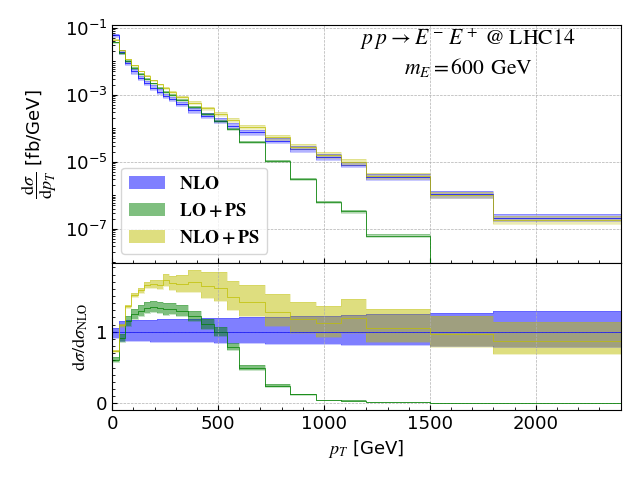}
    \includegraphics[width=0.48\textwidth]{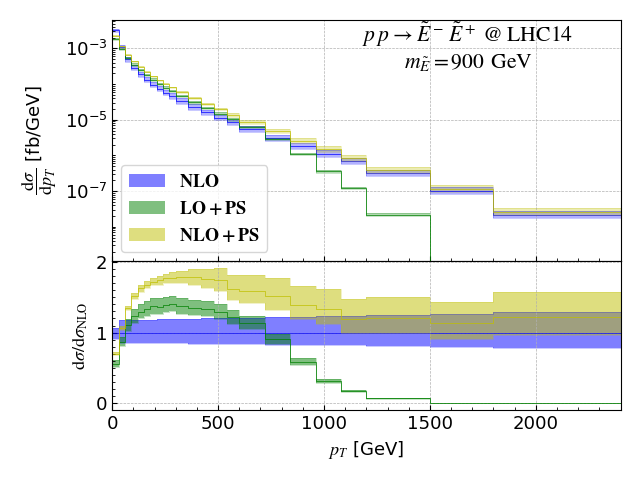}
    \includegraphics[width=0.48\textwidth]{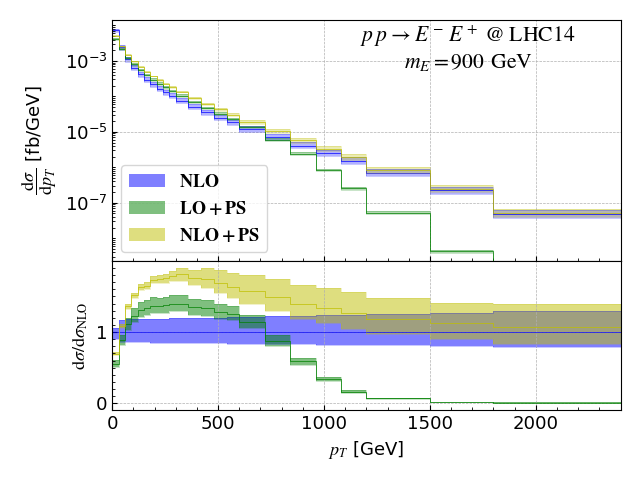}
    \includegraphics[width=0.48\textwidth]{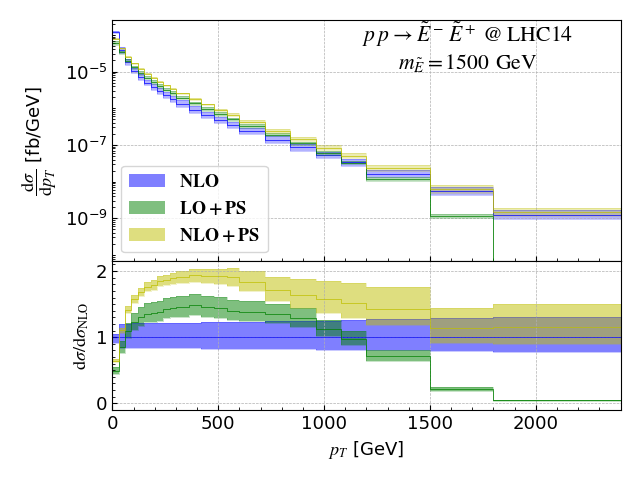}
    \includegraphics[width=0.48\textwidth]{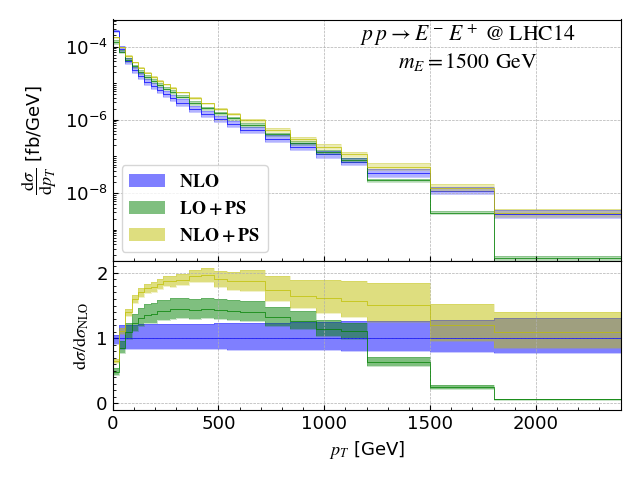}
    \caption{Transverse momentum spectra for the processes~\eqref{eq:process_vll}, together with the associated scale uncertainties (upper inset in each panel). We show fixed-order results at NLO (blue), as well as LO+PS (green) and NLO+PS (olive) predictions. We consider lepton masses of 600 GeV (top row), 900~GeV (middle row) and 1.5~TeV (bottom row), and we additionally present the corresponding bin-by-bin ratios to the NLO spectra with the associated scale uncertainties (lower inset in each panel).}\label{fig:dsdpt_vll}
\end{figure*}

In order to better assess the impact of threshold resummation, we display in the lower insets of figures~\ref{fig:dsdM_vll} and \ref{fig:dsdM_typeiii} bin-by-bin ratios of the NLO, NLO+NLL and NLO+NNLL rates to the most precise NLO+NNLL predictions evaluated with central scale choices. The bands represent again the associated scale uncertainties. We observe that the increase of the differential NLO cross section induced by NLL or NNLL resummation (or equivalently, the decrease of the NLO cross sections, shown in blue, relative to the most precise NLO+NNLL predictions) depends on the invariant mass of the di-lepton system $M$, and therefore indirectly on the mass of the lepton species produced that fix the kinematic production threshold $M_0$. Preferred configurations hence naturally target larger $z$ values closer to 1 for heavy lepton production than for light lepton production. Resummation effects are therefore expected to be more important for heavy leptons, when considering $M$ values lying at a given relative distance from $M_0$. Again, we observe that NLO+(N)NLL results are generally outside the NLO error bands.

Nevertheless, the shape of the spectrum is stabilised after including threshold resummation at NLL (yellow). NNLL resummation (purple) only yields a mild increase of the rate by less than 1\%, which is largely independent of the di-lepton invariant mass $M$. This shows that a good perturbative convergence has been achieved at NLO+NNLL. Moreover, NLO+NNLL predictions are crucial to reduce the scale uncertainties to be less than 0.5\%, hence motivating using more precise predictions for BSM signals when available.

\subsection{Transverse momentum spectra}\label{sec:dsdpt}

\begin{figure}
    \centering
    \includegraphics[width=0.48\textwidth]{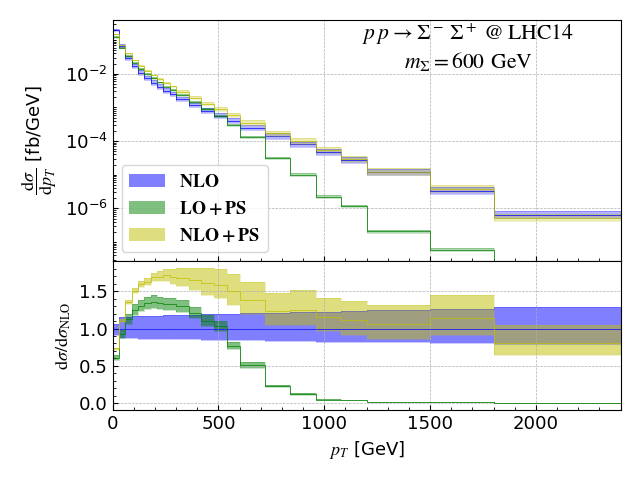}
    \includegraphics[width=0.48\textwidth]{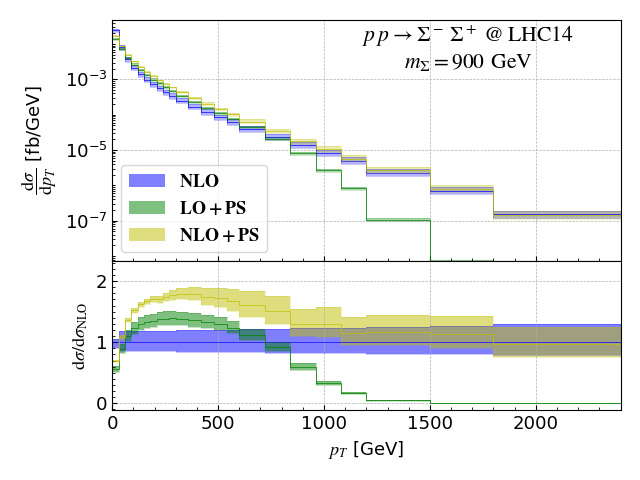}
    \includegraphics[width=0.48\textwidth]{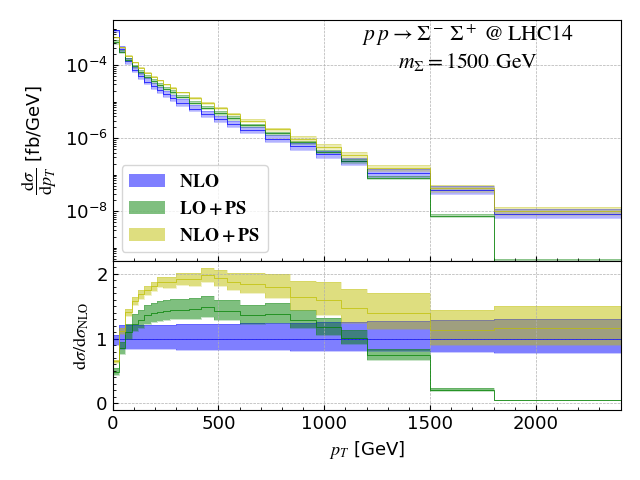}
    \caption{Same as figure~\ref{fig:dsdpt_vll} but for Type-III seesaw leptons.}
    \label{fig:dsdpt_typeiii}
\end{figure}

We now turn to the study of the distribution in the transverse momentum ($p_T$) of the lepton pair, which is a typical observable that shows that the inclusion of PS is essential. Although the \mgamc\ framework practically allows for investigations of any observable, we only focus, for the sake of an example, on this distribution in the $p_T$ of the di-lepton system. In figures~\ref{fig:dsdpt_vll} and \ref{fig:dsdpt_typeiii}, we present ${\rm d}\sigma/{\rm d}p_T$ distributions at NLO (blue), LO+PS (green) and NLO+PS (olive), whereas the LO distributions are trivially located at $p_T=0$. As in section~\ref{sec:dsdm}, we choose three benchmark scenarios featuring extra leptons with masses of 600~GeV, 900~GeV and 1.5~TeV respectively. The different $p_T$ spectra are shown in the upper insets of the figures, whereas the lower insets display their bin-by-bin ratios to the (fixed-order) NLO predictions. 

While NLO predictions (blue) in principle diverge at small $p_T$ due to uncancelled soft and/or collinear singularities originating from real emission, the integration of the differential cross section within a given bin regularises this divergence, the bin-by-bin results shown in the figures being normalised by the bin size. We therefore observe a finite cross section with a pronounced maximum in the low $p_T$ region, with $p_T\lesssim 30$~GeV (which corresponds to the first bin shown in the plots). The cross sections in the small $p_T$ regime are therefore expected to be better described by matching fixed-order predictions with PS. Showering, as well as the non-perturbative intrinsic transverse momentum $k_T$ of the constituents of the protons, should additionally distort the shapes of the spectra in the low $p_T$ regime. NLO+PS predictions are hence smaller than NLO ones at small $p_T\lesssim 30$~GeV, before becoming considerably greater at intermediate transverse momenta ($p_T \in [60, 500]$~GeV). At larger $p_T$ values, QCD radiation encoded in the real matrix elements dominates, so that NLO+PS predictions agree with NLO ones. 

This last effect can be even more evidenced by studying the LO+PS curves (green). Non-zero $p_T$ values are here purely arising from shower effects, which gives rise to a too soft spectrum in the high-$p_T$ regime. In conclusion, NLO+PS should be the best for describing such an observable, while NLO (LO+PS) predictions fail at low (high) $p_T$. Finally, we can note that multiplying LO+PS predictions by an overall $K$-factor, as traditionally done in many experimental and phenomenological studies, is unjustified in the aim of an accurate signal description, and will even yield qualitatively wrong results at high $p_T$.

\section{Conclusion}\label{sec:conclusion}
Numerous extensions of the Standard Model feature additional leptons that carry a variety of different electric charges. They are consequently actively searched for at collider experiments. In this work, we have studied the production of these extra leptons in effective frameworks representative of the TeV-scale phenomenology of several models featuring additional leptons. By providing {\sc FeynRules} implementations and the associated UFO libraries for Type-III seesaw models and new physics scenarios with VLLs, we complete the set of publicly available models suitable for calculations relevant for the production of new leptons at colliders beyond the LO or LO+PS accuracy. With these public models, VLL and Type-III lepton production can now be simulated at NLO+PS accuracy, as was already the case for other neutrino mass models or left-right-symmetric scenarios which both involve new non-coloured fermions. The corresponding model files can be downloaded from \url{https://feynrules.irmp.ucl.ac.be/wiki/NLOModels}.

We have reported the most precise calculations of total rates to date for the production of a pair of Type-III leptons or VLLs lying in the trivial or fundamental representations of $SU(2)_L$. Our predictions include both NLO QCD corrections and threshold resummation effects at NNLL. Higher-order QCD effects increase the production rates by 25\%--30\%, the exact value depending on the scenario and the extra lepton mass, and scale uncertainties are reduced below 1\%. We have additionally investigated the impact of these corrections on the distributions in the invariant mass of the produced heavy lepton system and observed a notable increase in the differential rates and a significant reduction of the scale uncertainties. 

Finally, we have made use of the designed UFO models to highlight the joint impact of NLO corrections and PS matching on an example of observable relevant for existing searches for extra leptons. We have chosen the distribution in the transverse momentum of the di-lepton system. We have illustrated how NLO+PS improves fixed-order NLO computations at low $p_T$, and provides a better modelling of the physical spectra at intermediate $p_T$ values below 1~TeV. These calculations have been achieved fully automatically, in the \mgamc framework, in a setup similar to that used in ATLAS and CMS studies as well as in many existing phenomenological explorations. Upgrades of existing studies and searches to include NLO corrections matched with PS should therefore be straightforward.

\section*{Acknowledgments}
The authors are grateful to M.~Cirelli and G.~Soyez for motivating discussions during the course of this project. We acknowledge support from the European Union’s Horizon 2020 research and innovation programme (grant agreement No.~824093, STRONG-2020, EU Virtual Access ``NLOAccess''), the European Research Council (ERC grant agreement ID 101041109, ``BOSON''), the French ANR (grants ANR-20-CE31-0015, ``PrecisOnium'', and ANR-21-CE31-0013, ``DMwithLLPatLHC''), and the CNRS IEA (grant No.~205210, ``GlueGraph").

%%%%%%%%%%%%%%%%%%%%%%%%%%%%%%%%%%%%%%%%

% \clearpage
\appendix
\input{appendix}
\bibliographystyle{utphys}
\bibliography{bibliography}
\end{document}

%% file: appendix.tex
\section{Resummation coefficients}\label{app:resummationcoefficiebts}
In this section, we provide analytical results for the resummation coefficients appearing in~\eqref{DeltaN} in the case of a Drell-Yan-like process. Those comprise the process-dependent term $\tilde{g}_{0,q\bar q}$, as well as the universal coefficients of the exponent $g_{k,q\bar{q}}$ with $k>0$. The latter are identical to those given in appendix~A of \cite{Ajjath:2019neu} (the $\lambda$ parameter of \cite{Ajjath:2019neu} being replaced by $\omega$). We then provide below the expression for the process-dependent coefficient $\tilde{g}_{0,q\bar q}(M^2,\mu_F^2,\mu_R^2)$ that can be expanded in $a_s$ as
\begin{equation}
    \tilde{g}_{0,q\bar q}(M^2,\mu_F^2,\mu_R^2) = \frac{{\rm d}\hat\sigma^{(0)}_{q\bar q}}{{\rm d}\ln{M^2}}\Big(1 + \sum_{k=1}{a_s^k(\mu_R^2) \tilde{g}_{0,q\bar q}^{(k)}}\Big)\,,
\end{equation}
where $\frac{{\rm d}\hat\sigma^{(0)}_{q\bar q}}{{\rm d}\ln{M^2}}$ is the Born partonic cross section differentiating with respect to $\ln{M^2}$. The $\tilde{g}_{0,q\bar q}^{(1)}$ and $\tilde{g}_{0,q\bar q}^{(2)}$ coefficients in QCD (for $n_q=5$) relevant for resummation at NNLL explicitly read
\begin{equation}\begin{split}
  \tilde{g}_{0,q\bar q}^{(1)} =&\ \frac{-64}{3} +  \frac{64}{3}  \zeta_2 - 8 L_{fr} + 8 L_{qr}\,,  \\
  \tilde{g}_{0,q\bar q}^{(2)} =&\ \frac{-1291}{9} + \frac{64 \zeta_2}{9} + \frac{368 \zeta_2^2}{3}  + \frac{4528 \zeta_3}{27}\\
    &\ + \frac{188 L_{fr}^2}{3} + \frac{4 L_{qr}^2}{3} \\
    &\ + L_{fr}  \bigg( \frac{1324}{9} - \frac{1888 \zeta_2}{9} + \frac{32 \zeta_3}{3} \bigg) \\
    &\  + L_{qr}  \bigg( \frac{148}{9} - 64 L_{fr} + \frac{416 \zeta_2}{9} - \frac{32 \zeta_3}{3} \bigg)\,,
\end{split}\end{equation}
with $L_{qr} = \ln \frac{M^2}{\mu_R^2}$, $L_{fr} = \ln \frac{\mu_F^2}{\mu_R^2}$ and $\zeta_n$ being the Riemann zeta function $\zeta(n)$.

In the processes \eqref{eq:process_vll} and \eqref{eq:process_typeiii}, the di-lepton system is produced either through an $s$-channel virtual photon exchange, or through an $s$-channel $Z$-boson exchange. In other words, the Born partonic cross section can be split into three pieces, namely the square of photon-exchange diagram, that of the $Z$-exchange diagram, and the interference between the two diagrams. Mathematically, this can be written as
\begin{align}\label{gamZ}
    \hat \sigma^{(0)}_{q\bar q} = \hat \sigma^{(0),\gamma}_{q\bar q} + \hat \sigma^{(0),Z}_{q\bar q} +  \hat \sigma^{(0),{\rm int}}_{q\bar q}\,.
\end{align}

The photon exchange contribution depends on the electric charge of the initial-state quarks $Q_q$, and on that of the final-state leptons $Q_{\ell}$,
\begin{align}\label{siggamgam}
    \hat \sigma^{(0),\gamma}_{q\bar q} = Q_q^2 Q_{\ell}^2 \frac{4 \pi \alpha^2  }{9 M^2} \sqrt{ 1 - \frac{4 m_{\ell}^2}{ M^2}} 
    \left(1 +  \frac{2 m_{\ell}^2}{ M^2}   \right)\,.
\end{align}
In this expression, $m_\ell$ is the mass of the produced lepton, and it is thus respectively given by $m_{\tilde E}$, $m_E$ and $m_{\Sigma}$ in the three processes shown in \eqref{eq:process_vll} and \eqref{eq:process_typeiii}. As all exotic leptons produced in these processes satisfy $Q_\ell=-1$, $\hat \sigma^{(0),\gamma}_{q\bar q}$ is identical in all three cases.

The $Z$-boson exchange contribution and the interference term read
\begin{eqnarray}
    \hat \sigma^{(0),Z}_{q\bar q} &=& f_\ell^2\frac{ (V_q^2 + A_q^2)}{4} 
      \dfrac{M^4}{(M^2 - m_Z^2)^2 + \Gamma_Z^2 m_Z^2} \frac{\hat \sigma_{q\bar q}^{(0),\gamma}}{Q_q^2}\,,\nonumber
      \\  
      \hat \sigma^{(0),{\rm int}}_{q\bar q} &=&f_\ell
       Q_q V_q
    \dfrac{M^2( M^2 -m_Z^2)}{(M^2 - m_Z^2)^2 + \Gamma_Z^2 m_Z^2} \frac{\hat \sigma_{q\bar q}^{(0),\gamma}}{Q_q^2}\,,
\end{eqnarray}
where $V_q=I_q-2s_{\sss W}^2Q_q$ and $A_q=I_q$ represent the vector and axial couplings between the initial quarks and the $Z$ boson respectively. Here, the parameters $c_{\sss W}$ and $s_{\sss W}$ are the cosine and sine of the electroweak mixing angle, and $I_q$ is the weak isospin quantum number of the quark $q$, that is thus equal to $1/2$ for up-type quarks and $-1/2$ for down-type quarks. In addition, $\Gamma_Z$ denotes the width of the $Z$ boson, and the prefactor $f_\ell$ depends on the quantum numbers of the lepton $\ell$ produced. It is thus given by%
\renewcommand{\arraystretch}{1.7}%
\begin{equation}
  f_\ell=\left\{\begin{array}{ll} 
    -c_{\sss W}^{-2} &\qquad \text{for}\quad \ell=\tilde{E}\,, \\ \frac{c_{\sss W}^2-s_{\sss W}^2}{2s_{\sss W}^2c_{\sss W}^2} &\qquad \text{for}\quad  \ell=E\,, \\
    s_{\sss W}^{-2} &\qquad \text{for}\quad  \ell=\Sigma\,.\\ \end{array}\right.
\end{equation}